\documentstyle[12pt]{article}

\begin{document}

\title{The Dirac particle on central backgrounds and the anti-de Sitter 
       oscillator}

\author{Ion I. Cot\u aescu\\ {\it The West University of Timi\c soara,}\\{\it V. 
Parvan Ave. 4, RO-1900 Timi\c soara}}

\maketitle  

\begin{abstract}
It is shown that, for spherically symmetric static backgrounds, a simple 
reduced Dirac equation can be obtained by using  the Cartesian tetrad gauge in 
Cartesian holonomic coordinates. This equation is  manifestly covariant 
under rotations so that  the spherical coordinates can be separated in terms of 
angular spinors  like in  special relativity, obtaining a pair of radial 
equations and a specific form of the radial scalar product. As an example, we 
analytically solve  the anti-de Sitter oscillator giving the formula of the 
energy levels and the  form of the corresponding eigenspinors. 

%Pacs: 04.62.+v
\end{abstract}
\

\section{Introduction}
\

In the gauge field theory \cite{G} on curved space-time the physical meaning 
does not depend on the choice of the holonomic (natural) frame, or on the gauge 
of the tetrad field which defines the local ones \cite{W,MTW}. However, from 
the observer's point of view, these frames are not completely equivalent since 
the concrete space-time behavior depends on their choice. In general, this is 
studied with the help of  geometric models, which play here the role of  
kinetics. In the last years, many quantum models, involving Klein-Gordon or 
Dirac test particles on given backgrounds, have been worked out with the hope 
to find  analytic solutions. The main  results are the quantum modes of the 
Klein-Gordon particle on anti-de Sitter static backgrounds \cite{N} or on 
several deformations of them \cite{C}, as well as the solutions of the Dirac 
equation on space-times with particular metrics of special interest 
\cite{SOL, VIL}.     

An important case is that of the Dirac equation on spherically 
symmetric (central) static charts which have the global symmetry of the 
group $T(1)\otimes SO(3)$, of  time translations and  rotations of the 
Cartesian space coordinates. There is a gauge in which the tetrad field in 
spherical coordinates has only diagonal components  and another one where the 
axes of the local frame, defined by the tetrad field, are parallel with 
those of the Cartesian natural frame. Usually these are referred as the 
diagonal tetrad gauge and Cartesian gauge, respectively  \cite{BW}. 
In general, for deriving the Dirac equation one prefers the diagonal tetrad 
gauge where the result is obtained directly in spherical coordinates \cite{D}. 
Moreover, when one use the Cartesian gauge this is written also in spherical 
coordinates \cite{BW,VIL}. Despite of the obvious advantages of these 
coordinates we believe that the study of the Dirac equation in  Cartesian gauge 
and Cartesian natural coordinates is also interesting since, in this context,
the whole theory is manifestly covariant under  $T(1)\otimes SO(3)$ group.  
Then  the energy and  angular momentum are conserved like in special relativity 
from which we can take over the method of separation of variables.

Here we present this approach for  arbitrary central static metrics. It is 
shown that when both the natural and local frames are Cartesian, we can put 
the Dirac equation in a simple form by using an appropriate transformation of 
the spinor field \cite{V}. It results a reduced Dirac equation in Cartesian 
coordinates which is manifestly covariant 
under rotations. Therefore, the separation of  variables in spherical 
coordinates can be done  in terms of the angular momentum eigenspinors, like in 
special relativity \cite{BJDR,TH}. We obtain the radial equations and the form 
of the radial scalar product in the most general case of any central 
static metric, generalizing thus the well-known result of Brill and Wheeler 
\cite{BW}. Moreover, we show that in our approach we can  easily identify  the 
radial problems  with supersymmetry, which could be analytically solvable. 
The example we give is of the Dirac particle on a static chart of an anti-de 
Sitter background (i.e. the anti-de Sitter oscillator)  for which we determine 
the quantum modes.  
 
We start in the second section with a short review of the main notations and 
formulas. In Sec.3  we define the Cartesian gauge in Cartesian natural 
coordinates and we obtain the reduced Dirac equation. The next section is 
devoted to the separation of  variables in spherical coordinates which 
allows us to define an independent radial problem, while in Sec.5 we 
discuss the cases when this has  supersymmetry. The hidden supersymmetry of 
the anti-de Sitter oscillator is pointed out in Sec.6, where we present its 
complete  solution  giving the formula of the  energy levels and 
the form of the  energy eigenspinors up to normalization factors.

\section{Preliminaries}
\

Let us consider a chart where we have introduced the natural  
frame of the coordinates $x^{\mu}, \mu=0,1,2,3$.  We denote by $e_{\hat\mu}(x)$ 
the tetrad fields which define the local frames and by $\hat e^{\hat\mu}(x)$ 
that of the corresponding coframes. These have the usual orthonormalization 
properties   
\begin{equation}
e_{\hat\mu}\cdot e_{\hat\nu}=\eta_{\hat\mu \hat\nu}, \quad
\hat e^{\hat\mu}\cdot \hat e^{\hat\nu}=\eta^{\hat\mu \hat\nu}, \quad 
\hat e^{\hat\mu}\cdot e_{\hat\nu}=\delta^{\hat\mu}_{\hat\nu}, 
\end{equation}
where $\eta=$diag$(1,-1,-1,-1)$ is the Minkowski metric. The 1-forms of the local 
frames,  $d\hat x^{\hat\mu}=\hat e_{\nu}^{\hat\mu}dx^{\nu}$, allow one to write 
the line element  
\begin{equation}\label{(met)}
ds^{2}=\eta_{\hat\mu \hat\nu}d\hat x^{\hat\mu}d\hat x^{\hat\nu}=
g_{\mu \nu}(x)dx^{\mu}dx^{\nu},
\end{equation}   
which defines the metric tensor $g_{\mu \nu}$ of the natural frame. This raises 
or lowers the Greek indices (ranging from 0 to 3) while for the hat Greek ones 
(with the same range) we have to use the Minkowski metric, 
$\eta_{\hat\mu \hat\nu}$. The derivatives  
$\hat\partial_{\hat\nu}=e^{\mu}_{\hat\nu}\partial_{\mu}$ satisfy the 
commutation rules  
\begin{equation}
[\hat\partial_{\hat\mu},\hat\partial_{\hat\nu}]
=e_{\hat\mu}^{\alpha} e_{\hat\nu}^{\beta}(\hat e^{\hat\sigma}_{\alpha,\beta}-
\hat e^{\hat\sigma}_{\beta,\alpha})\hat\partial_{\hat\sigma}
=C_{\hat\mu \hat\nu 
\cdot}^{~\cdot \cdot \hat\sigma}\hat\partial_{\hat\sigma}
\end{equation}
defining the Cartan coefficients which help us to write the conecttion 
components in the local frames as   
\begin{equation}
\hat\Gamma^{\hat\sigma}_{\hat\mu \hat\nu}=e_{\hat\mu}^{\alpha}
e_{\hat\nu}^{\beta}
(\hat e^{\hat\sigma}_{\beta, \alpha}+\hat e_{\gamma}^{\hat\sigma}
\Gamma^{\gamma}_{\alpha \beta})=
\frac{1}{2}\eta^{\hat\sigma \hat\lambda}(C_{\hat\mu \hat\nu \hat\lambda}+
C_{\hat\lambda \hat\mu \hat\nu}+C_{\hat\lambda \hat\nu \hat\mu})
\end{equation}
while the notation  $\Gamma^{\gamma}_{\alpha \beta}$ stands for the usual 
Christoffel symbols.

Let $\psi$ be a Dirac free field of  mass $M$, defined on the   
space domain $D$.  In natural units, $\hbar=c=1$, its gauge invariant action 
\cite{BD} is     
\begin{equation}\label{(action)}
{\cal S}[\psi]=\int_{D} d^{4}x\sqrt{-g}\left\{
\frac{i}{2}[\bar{\psi}\gamma^{\hat\alpha}D_{\hat\alpha}\psi-
(\overline{D_{\hat\alpha}\psi})\gamma^{\hat\alpha}\psi] - 
M\bar{\psi}\psi\right\}
\end{equation}
where 
\begin{equation}
D_{\hat\alpha}=\hat\partial_{\hat\alpha}+\frac{i}{2}S^{\hat\beta \cdot}_{\cdot 
\hat\gamma}\hat\Gamma^{\hat\gamma}_{\hat\alpha \hat\beta}
\end{equation}
are the covariant derivatives of the spinor field  
and $g=\det(g_{\mu\nu})$. 
The Dirac matrices,  $\gamma^{\hat\alpha}$, and the generators of the reducible 
spinor representation of the $SL(2,C)$ group, $S^{\hat\alpha \hat\beta}$, 
satisfy
\begin{eqnarray}
&&\{ \gamma^{\hat\alpha}, \gamma^{\hat\beta} \}=2\eta^{\hat\alpha \hat\beta}, 
\qquad [\gamma^{\hat\alpha}, \gamma^{\hat\beta} ]=-4iS^{\hat\alpha \hat\beta},\\ 
&&~~~~~~~[ S^{\hat\alpha \hat\beta}, \gamma^{\hat\mu}]=
i(\eta^{\hat\beta \hat\mu}\gamma^{\hat\alpha}-
\eta^{\hat\alpha \hat\mu}\gamma^{\hat\beta}).
\end{eqnarray}
Thereby it results that the field equation,
\begin{equation}\label{(d)}
i\gamma^{\hat\alpha}D_{\hat\alpha}\psi - M\psi=0,
\end{equation}
derived from  (\ref{(action)})  can be written as 
\begin{equation}\label{(dd)}
i\gamma^{\hat\alpha}e_{\hat\alpha}^{\mu}\partial_{\mu}\psi - M\psi
+ \frac{i}{2} \frac{1}{\sqrt{-g}}\partial_{\mu}(\sqrt{-g}e_{\hat\alpha}^{\mu})
\gamma^{\hat\alpha}\psi
-\frac{1}{4}
\{\gamma^{\hat\alpha}, S^{\hat\beta \cdot}_{\cdot \hat\gamma} \}
\hat\Gamma^{\hat\gamma}_{\hat\alpha \hat\beta}\psi =0.
\end{equation}
On the other hand, from the conservation of the electric charge, we can deduce 
that when $e^{0}_{i}=0$,  $i=1,2,3$, then the time-independent relativistic 
scalar product of two spinors is \cite{BD}
\begin{equation}\label{(sp)}
(\psi,\psi')=\int_{D}d^{3}x\mu(x)\bar\psi(x)\gamma^{0}\psi'(x), \quad 
\end{equation}
where 
\begin{equation}\label{(weight)}
\mu(x)=\sqrt{-g(x)}e_{0}^{0}(x)
\end{equation}
is the specific weight function of the Dirac field.

\section{The reduced Dirac equation}
\

Our aim  is to discuss here only the case of  the  charts with the global 
symmetry of the $T(1)\otimes SO(3)$ group. These have natural frames of the 
Cartesian coordinates   $x^{0}=t$ and  $x^i$, $i=1,2,3$, in which  the metric 
tensor is time-independent and manifestly covariant  under the  rotations 
$R\in SO(3)$ of the space coordinates,
\begin{equation}\label{(rot)}  
x^{\mu}\to x'^{\mu}=(Rx)^{\mu} \qquad
(t'=t,\quad  x'^{i}= R_{ij}x^{j}).
\end{equation}
The most general form of a such a metric is given by the line element     
\begin{equation}\label{(metr)} 
ds^{2}=g_{\mu\nu}(x)dx^{\mu}dx^{\nu}=A(r)dt^{2}-[B(r)\delta_{ij}+C(r)x^{i}x^{j}]
dx^{i}dx^{j}
\end{equation} 
where  $A$, $B$ and $C$  are arbitrary functions of the Euclidian norm of 
$\stackrel{\rightarrow}{x}$,  $r=\vert\stackrel{\rightarrow}{x}\vert$ (which is 
invariant under rotations).  In applications it is convenient to replace these 
functions by  new ones,  $u$,  $v$ and $w$, such that
\begin{equation}\label{(ABC)}   
A=w^{2}, \quad B=\frac{w^2}{v^2}, \quad 
C=\frac{w^2}{r^2}\left( \frac{1}{u^2}-\frac{1}{v^2}\right).
\end{equation}
Then the metric appears as the conformal transformation of that simpler one 
having $w=1$.

Starting with a Cartesian natural frame we define the Cartesian gauge in which                                                            
the static tetrad field  transforms under the rotations (\ref{(rot)}) 
according to the rule
\begin{equation}\label{(tr)}
d\hat x ^{\hat\mu}\to d\hat x'^{\hat\mu}=\hat e^{\hat\mu}_{\alpha}(x')dx'^{\alpha}
=(Rd\hat x)^{\hat\mu}.
\end{equation}
In the case of the metric (\ref{(metr)}) the simplest choice of their 
components is
\begin{eqnarray}
\hat e^{0}_{0}&=&\hat a(r), \quad \hat e^{0}_{i}=\hat e^{i}_{0}=0, \quad
\hat e^{i}_{j}=\hat b(r)\delta_{ij}+\hat c(r) x^{i}x^{j},\label{(eee)}\\
e^{0}_{0}&=& a(r), \quad  e^{0}_{i}= e^{i}_{0}=0, \quad
e^{i}_{j}= b(r)\delta_{ij}+ c(r) x^{i}x^{j},\label{(eee1)}
\end{eqnarray}
where, according to (\ref{(met)}), (\ref{(metr)}) and (\ref{(ABC)}), we must 
have 
\begin{eqnarray}
\hat a&=&w, \quad \hat b=\frac{w}{v}, \quad \hat c=\frac{1}{r^2}
\left( \frac{w}{u}-\frac{w}{v}\right), \label{(abc)}\\
a&=& \frac{1}{w}, \quad  b=\frac{v}{w}, \quad  c=\frac{1}{r^2}
\left( \frac{u}{w}-\frac{v}{w}\right),\label{(abc1)}
\end{eqnarray}                                                                                                                                                                                                                                                 
      
while the weight function  (\ref{(weight)}) becomes
\begin{equation}\label{(mu)}
\mu
=\frac{1}{b^{2}(b+r^{2}c)}
=\frac{w^3}{uv^2}
\end{equation} 
since
\begin{equation}
\sqrt{-g}=B[A(B+r^{2}C)]^{1/2}=\frac{1}{ab^{2}(b+r^{2}c)}
=\frac{w^4}{uv^2}.
\end{equation}

From (\ref{(abc)}) and (\ref{(abc1)}) we see that the function $w$ must be 
positively defined in order to keep the same sense for the time axes of the 
natural and local frames. In addition, it is convenient to consider that the 
function $u$ is positively defined too. However, the function $v$ has an 
arbitrary sign. It can be represented as
\begin{equation}
v=\eta_{P} |v|
\end{equation}
where $\eta_{P}$ gives the relative parity. More precisely, when $\eta_{P}=1$  
then the space axes of the local frame are parallel with those of the natural 
frame, while if $\eta_{P}=-1$ these are antiparalel.

Now we have to replace the concrete form of the tetrad components in 
Eq.(\ref{(dd)}). First we eliminate its last term 
since it is known that this can not contribute when the metric is spherically 
symmetric. The argument is that 
$\{\gamma^{\hat\alpha},S^{\hat\beta \hat\gamma}\}=
\varepsilon^{\hat\alpha \hat\beta \hat\gamma \cdot}_{~~~ \hat\lambda}
\gamma^{5}\gamma^{\hat\lambda}$ (with $\varepsilon^{0123}=1$) is 
completely antisymmetric, while 
the Cartan coefficients resulted from (\ref{(eee)}) and  (\ref{(eee1)}) have 
no such kind of components. Furthermore,  in order to simplify the remaining 
equation, we introduce the {\it reduced} Dirac field, $\tilde\psi$, defined by  
\begin{equation}\label{(cfu)}
\psi(x)=\chi(r)\tilde\psi(x).
\end{equation}
where 
\begin{equation}\label{(chi)}
\chi=[\sqrt{-g}(b+r^{2}c)]^{-1/2}=b\sqrt{a}=vw^{-3/2}.     
\end{equation}
After this transformation  we obtain the reduced Dirac equation in Cartesian 
coordinates  and Cartesian tetrad gauge,
\begin{equation}\label{(red)}
i\{a(r)\gamma^{0}\partial_{t} +b(r)(\stackrel{\rightarrow}{\gamma}\cdot 
\stackrel{\rightarrow}{\partial})+
c(r)(\stackrel{\rightarrow}{\gamma}\cdot \stackrel{\rightarrow}{x})[1+
(\stackrel{\rightarrow}{x}\cdot \stackrel{\rightarrow}{\partial})]\}\tilde\psi(x)
-M\tilde\psi(x)=0.
\end{equation}
This is expressed only in terms of familiar three-dimensional scalar products 
and scalar functions so that it is manifestly covariant under 
rotations. Consequently, all the properties related to the conservation of 
the angular momentum, including the separation of  variables in spherical 
coordinates, will be similar as those of the usual  Dirac theory in the 
Minkowski flat space-time.  Moreover, we can verify that here the discrete 
transformations, $P$, $C$ and $T$, have the same significance as those of 
special relativity \cite{BJDR,TH}. Thus, for example, the charge conjugation transforms each 
particular solution of positive frequency of (\ref{(red)}) into the 
corresponding one of negative frequency.

\section{The radial problem}
\

The next step is to introduce the spherical coordinates, $r$, $\theta$, $\phi$,  
associated with  the space  coordinates of our natural Cartesian frame.  
Then from (\ref{(metr)}) and (\ref{(ABC)}) we obtain  the line element   
\begin{equation}\label{(muvw)}
ds^{2}=w^{2}\left[dt^{2}-\frac{dr^2}{u^2}-
\frac{r^2}{v^2}(d\theta^{2}+\sin^{2}\theta d\phi^{2})\right].
\end{equation}
Since this is static the energy, $E$, is conserved. On the other hand, the 
form of the reduced Dirac equation (\ref{(red)}) allows us to separate the 
spherical variables as  in the case of the central motion in flat space-time, 
by using the four-components angular spinors $\Phi^{\pm}_{m_{j}, \kappa_{j}}
(\theta, \phi)$ as given in Ref.\cite{TH}. These are orthogonal to each other 
being completely determined by the quantum number, $j$, of the  angular 
momentum, the quantum number, $m_{j}$, of its projection along the third axis 
(of the Cartesian frame) and the value of $\kappa_{j}=\pm (j+1/2)$. Based on 
these arguments, we consider the  particular  solution of positive frequency 
\begin{equation}\label{(psol)}
\tilde\psi_{E,j,m_{j},\kappa_{j}}(t,r,\theta,\phi)
=\frac{1}{r}[f^{(+)}(r)\Phi^{+}_{m_{j},\kappa_{j}}(\theta,\phi)
+f^{(-)}(r)\Phi^{-}_{m_{j},\kappa_{j}}(\theta,\phi)]e^{-iEt}.
\end{equation}
By  replacing it in (\ref{(red)}), after a little calculation, one finds 
that the radial functions $f^{(\pm)}$ must satisfy the radial equations
\begin{eqnarray}
\left[u(r)\frac{d}{dr}+v(r)\frac{\kappa_j}{r}\right]f^{(+)}(r)&=&[E+w(r)M
]f^{(-)}(r),\label{(e1)}\\
\left[- u(r)\frac{d}{dr}+v(r)\frac{\kappa_j}{r}\right]f^{(-)}(r)&=&[E-w(r)M
]f^{(+)}(r).\label{(e2)}
\end{eqnarray}
In practice, these can be written directly  starting with the line element put 
in the form (\ref{(muvw)}) from  which we have to  identify  the functions $u$, 
$v$, and $w$ we need.

The angular spinors are normalized such that the angular integral of the scalar 
product (\ref{(sp)}) does not contribute and, consequently, this reduces to the 
radial integral. By using (\ref{(cfu)}) and (\ref{(psol)}) we find that this is
\begin{equation}\label{(spp)}
(\tilde\psi_{1},\tilde\psi_{2})=\int_{D_{r}}\frac{dr}{u(r)}\{[f_{1}^{(+)}(r)]^{*}
f_{2}^{(+)}(r)+[f_{1}^{(-)}(r)]^{*}f_{2}^{(-)}(r)\}
\end{equation}
where $D_{r}$ is the radial domain corresponding to $D$. What  is remarkable 
here is that the weight function $\mu \chi^{2}=1/u$, resulted from (\ref{(mu)}) 
and (\ref{(chi)}), is just that we need in order to have 
$(u\partial_{r})^{+}=-u\partial_{r}$. 
This means that the operators of the left-hand side of the radial equations,
 are related between them through the Hermitian conjugation with respect to 
the scalar product (\ref{(spp)}). 

A direct consequence is that the operator 
\begin{equation}
H=\begin{array}{|cc|}
    Mw& -u\frac{\textstyle d}{\textstyle dr}+\kappa_{j}\frac{\textstyle v}
{\textstyle r}\\
       u\frac{\textstyle d}{\textstyle dr}+\kappa_{j}\frac{\textstyle v}
{\textstyle r}& -Mw
\end{array}
\end{equation}
is self-adjoint. This is the radial Hamiltonian, which allows one to write the 
Eqs.(\ref{(e1)}) and (\ref{(e2)}) as the eigenvalue problem   
\begin{equation}\label{(hfef)}
H{\cal F}=E{\cal F},
\end{equation}   
where the two-dimensional eigenvectors,  ${\cal F}=|f^{(+)}, f^{(-)}|^{T}$, have their own 
scalar product,
\begin{equation}\label{(spf)}
({\cal F}_{1},{\cal F}_{2})=\int_{D_{r}}\frac{dr}{u}
{\cal F}_{1}^{+}{\cal F}_{2},
\end{equation}
as it results from (\ref{(spp)}). Thus we have obtained an independent radial 
problem which must be solved in  each particular case separately by using 
appropriate methods.

\section{Supersymmetry in special frames}
\

First of all, we  look for  possible transformations which should simplify the 
radial equations. It is known that the transformations of the space coordinates 
of a natural frame with static metric do not change the quantum modes. The 
simplest ones are  the changes of the radial coordinate which allow us to 
choose  suitable frames with spherical symmetry. In our opinion, the best  
choice is that in which the radial coordinate is defined by
\begin{equation}\label{(spfr)}
r_{s}(r)=\int\frac{dr}{u(r)}+const
\end{equation}
so that  $r_{s}(0)=0$.  This will be called  the radial coordinate 
of the {\it special} frame. In the following we shall use only this frame by 
taking directly $u=1$ while the subscript $s$ will be omitted.     

Other transformations which could simplify the radial problem are the unitary 
transformations, ${\cal F}\to \hat {\cal F}=U{\cal F}$ and $H\to \hat 
H=UHU^{+}$. We shall take 
only those unitary matrices, $U$, which commute with the term of $H$ containing 
derivatives. It is clear that these are nothing else than  simple rotations 
of the plane $\{f^{(+)},f^{(-)}\}$. Furthermore, like in the Dirac theory in flat 
space-time \cite{TH}, we shall say that a radial problem has supersymmetry if 
there exist a  rotation of this kind such that the transformed Hamiltonian  
takes the form
\begin{equation}\label{(ssh)}
\hat H=\begin{array}{|cc|}
    \nu& -\frac{\textstyle d}{\textstyle dr}+W\\
       \frac{\textstyle d}{\textstyle dr}+W& -\nu
\end{array}
\end{equation}
where $\nu$ must be a constant and $W$ is the resulting superpotential 
\cite{SUP}. If the radial problem has this property, then the second order 
equations for the components $\hat f^{(+)}$ and $\hat f^{(-)}$ of 
$\hat{\cal F}$ can be obtained from $\hat H^{2}\hat{\cal F}=E^{2}\hat{\cal F}$. 
These equations,  
\begin{equation}\label{(fpm)}
\left(-\frac{d^{2}}{dr^2}+W(r)^{2}\mp \frac{dW(r)}{dr}+\nu^{2}\right)
\hat f^{(\pm)}(r)=E^{2}\hat f^{(\pm)}(r),
\end{equation}
represent the starting point for finding analytical solutions.

The simplest radial problems are those with manifest supersymmetry, for which  
the original radial Hamiltonian $H$ has the form (\ref{(ssh)}). 
These are generated by the metrics of the  central manifolds $R\times M_{3}$ 
which have $w=1$.  In the special frames these  are  determined only by 
the arbitrary function $v$  which gives the superpotential $W=\kappa_{j}v/r$.

A more complicated situation is when the metrics  are conformal 
transformations trough $w^{2}$ of the previous ones, with  functions $w$ 
of the form 
\begin{equation}
w=c_{1}+c_{2}\frac{v}{r}
\end{equation} 
where $c_{1}$ and $c_{2}$ are constants. In this case we need to use a suitable 
rotation $U$ in order to point out the supersymmetry.  These are  
problems with  hidden supersymmetry which are similar with that of the Dirac 
particle in external Coulomb field, known from  special relativity. 

However, an example in which the supersymmetry is much more hidden  will be 
discussed in the next section.

\section{The anti-de Sitter oscillator}
\

Let us consider a Dirac test particle on an anti-de Sitter  background,  
in the static chart of coordinates $(t,\hat r, \theta, \phi)$ where the 
metric is given  by the line element
\begin{equation}
ds^{2}=(1+\omega^{2}\hat r^{2})dt^{2}
-\frac{d\hat r^2}{1+\omega^{2}\hat r^{2}}
-\hat r^{2} (d\theta^{2}+\sin^{2}\theta~d\phi^{2}).
\end{equation}
In the special frame, $(t,r,\theta,\phi)$, defined according to (\ref{(spfr)}),
the radial coordinate is
\begin{equation}
r=\frac{1}{\omega}\arctan \omega \hat r
\end{equation}
and the line element becomes 
\begin{equation}\label{(le)}
ds^{2}=\csc^{2}\omega r \left[dt^{2}-dr^{2}-\frac{1}{\omega^{2}}
\sin^{2}\omega r~ (d\theta^{2}+\sin^{2}\theta~d\phi^{2})\right].
\end{equation} 
It defines a metric which covers the radial domain $D_{r}=[0,r_{e})$, bounded 
by the event horizon which is at $r_{e}=\pi/2\omega$. 

Working in the special frame we have $u=1$. The other two functions can 
be identified from (\ref{(le)}) as
\begin{equation}
w(r)=\csc \omega r, \quad v(r)=\omega r \sec \omega r.
\end{equation}
With their help and by using the notation $k=M/\omega$ (i.e. 
$Mc^{2}/\hbar\omega$ in usual units), we obtain the Hamiltonian
of the radial problem
\begin{equation}
H=\begin{array}{|cc|}
    \omega k\csc \omega r& -\frac{\textstyle d}{\textstyle dr}+
\omega\kappa_{j}\sec\omega r\\
       \frac{\textstyle d}{\textstyle dr}+\omega\kappa_{j}\sec\omega r
& -\omega k\csc \omega r
\end{array}.
\end{equation}
Its form suggests us to  transform  ${\cal F}$ into $\hat{\cal F}=U(r){\cal F}$  
by using the local rotation  
\begin{equation}\label{(uder)} 
U(r)=\begin{array}{|cc|}
    \cos \frac{\textstyle \omega r}{\textstyle 2}&-\sin 
\frac{\textstyle \omega r}{\textstyle 2}\\
    \sin \frac{\textstyle \omega r}{\textstyle 2}&\cos 
\frac{\textstyle \omega r}{\textstyle 2}
\end{array}.
\end{equation}
A little calculation shows us that  the transformed  Hamiltonian,   
\begin{equation}\label{(newh)}
\hat H =U(r)HU^{+}(r)-\frac{\omega}{2} 1_{2\times 2},
\end{equation}
which gives the eigenvalue problem
\begin{equation}\label{(trrp)}
\hat H \hat{\cal F}=\left(E-\frac{\omega}{2}\right)\hat{\cal F},
\end{equation} 
has supersymmetry since it  has  the requested specific form  (\ref{(ssh)}) 
with  $\nu=\omega(k-\kappa_{j})$  and the superpotential
\begin{equation}\label{(super)}
W(r)=\omega(k\tan\omega r + \kappa_{j}\cot \omega r). 
\end{equation}
Consequently, the components $\hat f^{(\pm)}$ of $\hat{\cal F}$ satisfy the 
second order equations 
\begin{equation}
\left(-\frac{d^2}{dr^2}+\omega^{2}\frac{k(k\mp 1)}{\cos^{2}\omega r}+
\omega^{2}\frac{\kappa_{j}(\kappa_{j}\pm 1)}{\sin^{2}\omega r}\right)
\hat f^{(\pm)}(r)=
\omega^{2}\epsilon^{2}\hat f^{(\pm)}(r)\label{(od1)},\\
\end{equation}
where  $\epsilon=E/\omega-1/2$. 

These equations are well-studied. Their solutions can be expressed in terms of 
hypergeometric functions \cite{AS} depending on the new variable 
$y=\sin^{2}\omega r$ as 
\begin{equation}\label{(gsol)}
\hat f^{(\pm)}(y)=N_{\pm}(1-y)^{p_{\pm}}y^{s_{\pm}}
F\left(s_{\pm}+p_{\pm}-\frac{\epsilon}{2},
s_{\pm}+p_{\pm}+\frac{\epsilon}{2}, 2s_{\pm}+\frac{1}{2}, y\right).
\end{equation}
They are considered on the domain $D_{y}=[0,1)$  corresponding 
to $D_{r}$ and  depend on the parameters  $p_{\pm}$ and $s_{\pm}$ which must 
accomplish  
\begin{eqnarray}\label{(kka)}
2s_{\pm}(2s_{\pm}-1)=\kappa_{j}(\kappa_{j}\pm 1)\\ 
2p_{\pm}(2p_{\pm}-1)=k(k\mp 1). 
\end{eqnarray}
Thus we have the general form of the solutions of the second order 
equations up to normalization factors, $N_{\pm}$. It remains to precise 
the values of the parameters $s_{\pm}$ and $p_{\pm}$ and the value of 
$N_{+}/N_{-}$ so that the functions $\hat f^{(\pm)}$ should be solutions of the 
transformed radial problem (\ref{(trrp)}), with  a good physical meaning.

The hypergeometric functions on the domain $[0,1)$ can  be either polynomials 
or analytical functions strongly divergent for $y\to 1$ which can not be 
interpreted as tempered distributions corresponding to continuous energy 
levels. Therefore, we have to look only for square integrable 
eigenfunctions of the discrete energy spectrum. These can be obtained 
by choosing  regular solutions in $y=0$ and $y=1$, with  $p_{\pm}\ge 0$ 
and $s_{\pm}\ge 0$, and by imposing the particle-like (with $\epsilon >0$) 
quantization conditions
\begin{equation}\label{(quant)}
\epsilon=2 (n_{\pm}+s_{\pm}+p_{\pm})
\end{equation}
which must be compatible, i.e.
\begin{equation}\label{(comp)}
n_{+}+s_{+}+p_{+}=n_{-}+s_{-}+p_{-}.
\end{equation}
Consequently, the solutions (\ref{(gsol)}) become 
\begin{equation}\label{(gsol1)}
\hat f^{(\pm)}(y)=N_{\pm}(1-y)^{p_{\pm}}y^{s_{\pm}}
F\left(-n_{\pm},
2s_{\pm}+2p_{\pm}+n_{\pm}, 2s_{\pm}+\frac{1}{2}, y\right)
\end{equation}
In the following we shall select the values of the parameters involved 
herein for each type of solution separately by using only one radial quantum 
number, $n_{r}$. Moreover, it is convenient to consider explicitly the value 
of the orbital angular momentum  quantum number, $l$, of the angular spinor 
$\Phi^{+}_{m_{j},\kappa_{j}}$, as an auxiliary quantum number.

Let us  take first $\kappa_{j}=-(j+1/2)=-l-1$. Then the positive solutions of 
the equations (\ref{(kka)}) are 
\begin{equation}
2s_{+}=l+1,\quad 2s_{-}=l+2,\quad 2p_{+}=k, \quad  2p_{-}=k+1,
\end{equation} 
while, according to (\ref{(comp)}), we must have 
\begin{equation}
n_{+}=n_{r}, \quad  n_{-}=n_{r}-1. 
\end{equation}
Furthermore, we  verify that, for these values of the parameters, 
the functions  (\ref{(gsol1)}) represent a solution of the transformed radial 
problem if and only if
\begin{equation}
\frac{N_{-}}{N_{+}}=-\frac{2n_{r}}{2l+1}
\end{equation} 
Thus we arrive at the final result in the special frame,  
\begin{eqnarray}
\hat f^{(+)}(r)&=& \left(l+\frac{1}{2}\right)\cos^{k}\omega r 
\sin^{l+1}\omega r\nonumber\\ 
&&\times F\left(-n_{r},n_{r}+k+l+1, l+\frac{3}{2}, \sin^{2}\omega r\right)
\label{(1)}\\
\hat f^{(-)}(r)&=&-n_{r} \cos^{k+1}\omega r \sin^{l+2}\omega r\nonumber\\ 
&&\times F\left(-n_{r}+1,n_{r}+k+l+2, l+\frac{5}{2}, \sin^{2}\omega r
\right)\nonumber
\end{eqnarray}
 
For $\kappa_{j}=j+1/2=l$ we  find 
\begin{eqnarray}
2s_{+}=l+1,\quad 2s_{-}=l,\quad 2p_{+}=k, \quad  2p_{-}=k+1, \\
n_{+}=n_{-}=n_{r}~~~~~~~~~~~~~~~~~~~~~~~~~
\end{eqnarray} 
and
\begin{equation}
\frac{N_{-}}{N_{+}}=\frac{2l+1}{2n_{r}+2k+1}
\end{equation} 
so that the solutions can be written as
\begin{eqnarray}
\hat f^{(+)}(r)&=& \left(n_{r}+k+\frac{1}{2}\right)\cos^{k}\omega r 
\sin^{l+1}\omega r\nonumber\\ 
&&\times F\left(-n_{r},n_{r}+k+l+1, l+\frac{3}{2}, \sin^{2}\omega r\right)\label{(4)}\\
\hat f^{(-)}(r)&=&\left(l+\frac{1}{2}\right) \cos^{k+1}\omega r \sin^{l}\omega r\nonumber\\ 
&&\times F\left(-n_{r},n_{r}+k+l+1, l+\frac{1}{2}, \sin^{2}\omega r\right)\nonumber
\end{eqnarray}

The energy levels result from (\ref{(quant)}). Bearing in mind that $\omega k=
M$ and $\omega\epsilon=E-\omega/2$, and by using the main quantum number 
$n=2n_{r}+l$ we obtain 
\begin{equation}\label{(enlev)}
E_{n}=M+\omega\left(n+\frac{3}{2}\right).
\end{equation} 
These levels are degenerated. For a given $n$ our auxiliary quantum number $l$ 
takes either all the odd values from $1$ to $n$ if $n$ is odd, or the even 
values from $0$ to $n$ if $n$ is even. In both cases  we have $j=l\pm 1/2$ 
for each $l$ which means that $j=1/2,3/2,...,n+1/2$. The selection rule for 
$\kappa_{j}$ is more complicated since it is determined by both the 
quantum numbers  $n$ and $j$. 
If $n$ is even then the even $\kappa_{j}$ are positive while the odd 
$\kappa_{j}$ are negative. For odd $n$ we are in the opposite situation, with 
odd positive or even negative values of $\kappa_{j}$. Thus it is clear that for 
each given pair $(n,j)$ we have only one value of $\kappa_{j}$. 
This means that the degree of degeneracy of the level $E_{n}$ is $n+1$.

On the other hand, if we know the values of $n$, $j$ and $\kappa_{j}$, then 
we can find those of the quantum numbers $n_{r}$ and $l$ of 
the solutions (\ref{(1)}) and (\ref{(4)}). For this reason these will be 
denoted by $\hat f^{(\pm)}_{n,j}$. With their help we can write  
the components of ${\cal F}$  by using the inverse of (\ref{(uder)}). Finally, 
from (\ref{(psol)}), (\ref{(cfu)}) and (\ref{(chi)}) we restore the  
form of the positive frequency  energy eigenspinors in the special frame, up to a 
normalization factor, $N_{n,j}$,  
\begin{eqnarray}
&&u_{n,j,m_{j}}(r,\theta,\phi)=\nonumber\\
&&=N_{n,j}\sec \omega r \cos^{3/2}\omega r \left[
\left(\cos\frac{\omega r}{2}\hat f^{(+)}_{n,j}(r)+\sin\frac{\omega r}{2}
\hat f^{(-)}_{n,j}(r)\right)\Phi^{+}_{m_{j},\kappa_{j}}(\theta,\phi)\right.\nonumber\\    
&&+\left.\left(-\sin\frac{\omega r}{2}\hat f^{(+)}_{n,j}(r)+\cos\frac{\omega r}{2}
\hat f^{(-)}_{n,j}(r)\right)\Phi^{-}_{m_{j},\kappa_{j}}(\theta,\phi)\right].
\end{eqnarray}
The negative frequency eigenspinors can be derived directly by using the charge 
conjugation \cite{BJDR}. These are  
\begin{equation}
v_{n,j,m_{j}}=(u_{n,j,m_{j}})^{c}\equiv C (\bar u_{n,j,m_{j}})^{T} 
\end{equation}
where $C=i\gamma^{2}\gamma^{0}$. Thus the problem of the quantum modes of the 
anti-de Sitter oscillator is completely solved.

\section{Comments}
\

The above presented example shows us  that our method based on the Cartesian 
tetrad gauge in Cartesian coordinates  has similar features with that used to 
solve  the central motion in flat space-time. We can say that, in some sense, 
the pair of the natural and local frames we have choose plays the same role 
as  the rest frames from special relativity. This allowed us to separate the 
spherical variables in terms of the angular spinors such that  all the 
constants involved in the separation of variables get good physical meaning. 
On the other hand, the complete formulation of the radial problem is useful in 
applications since it  includes the radial scalar product which help us to 
identify the radial wave functions corresponding to the discrete or continuous 
energy spectra.

By using this method we have obtained the quantum modes  
of the anti-de Sitter oscillator. In our opinion, this is the first step to the 
quantum  theory of the Dirac field on anti-de Sitter backgrounds. It follows to 
calculate the normalization factors, to derive the main properties of the 
spinors $u_{n,j,m_{j}}$ and $v_{n,j,m_{j}}$, and to introduce suitable creation 
and anihilations operators. Thus we hope to obtain in near future the  quantum 
theory of a free Dirac field (in the sense of general relativity) with 
countable energy spectrum.


\begin{thebibliography}{20}

\bibitem{G} 
R. Utiyama, Phys. Rev. {\bf 101}, 1597 (1956);
T. W. B. Kibble, J. Math. Phys. {\bf 2}, 212 (1961)
 
\bibitem{W}
S. Weinberg, {\it Gravitation and Cosmology: Principles and Applications of 
the General Theory of Relativity}, Wiley, New York, 1972


\bibitem{MTW}
C. M. Misner, K. S. Thorne and J. A. Wheeler, {\it Gravitation}, W. H. Freeman 
\& Co., San Francisco, 1973 


\bibitem{N}
D. J. Navarro and J. Navarro-Salas, J. Math. Phys. {\bf 37}, 6006 (1996)


\bibitem{C}
I. I. Cot\u aescu, Mod. Phys. Lett. A {\bf 12}, 685 (1997) 

\bibitem{SOL}
V. S. Otchik, Class. Quant. Grav. {\bf 2}, 539 (1985);
L. P. Chimento and M. S. Mollerach, Phys. Rev. {\bf D34},3698 (1986);
Phys. Lett. A {\bf 121}, 7 (1987);
M. A. Costagnino, C. D. El Hasi, F. D. Mozzitelli and J. P. Paz, 
Phys. Lett. A {\bf 128}, 25 (1988)


\bibitem{VIL}
V. M. Vilalba and U. Percoco, J. Math. Phys. {\bf 31}, 715 (1990);
G. V. Shishkin, Class. Quant. Grav. {\bf 8}, 175 (1991);
G. V. Shishkin and V. M. Vilalba, J. Math. Phys. {\bf 30}, 2132 (1989);
J. Math. Phys. {\bf 33}, 2093 (1992)



\bibitem{BW} 
D. R. Brill and J. A. Wheeler, Rev. Mod. Phys. {\bf 29}, 465 (1957);  

\bibitem{D}
D. R. Brill and J. A. Cohen, J. Math. Phys. {\bf 7}, 238 (1966); 
J. Klauder and J. A. Wheeler, Rev. Mod. Phys. {\bf 29}, 516 (1957); 
T. M. Davis and J. R. Ray, J. Math. Phys. {\bf 16}, 75 (1975),
Phys. Rev.   {\bf D9}, 331 (1974),
J. Math. Phys. {\bf16}, 80 (1975); 
K. D. Kriori and H. Kakati, GRG {\bf 20}, 1237 (1995);
J. C. Huang, N. O. Santos and Kleber, Class. Quantum Grav. 
{\bf 12}, 1245 (1995);
I. D. Soares and J. Tiomno, Phys. Rev.  {\bf D54}, 2808 (1996);
C. G. De Oliveira and J. Tiomno, Il Nouvo Cimento {\bf 24}, 672 (1962);
B. D. B. Figueredo, I. D. Soares and Tiomno, Class. Quantum Grav. 
{\bf 9}, 1593 (1992);
Hammond R., Class. Quantum Grav. {\bf 12}, 279 (1995);
P. Baekler, M. Setz, V. Winkelmann, Class. Quantum Grav. {\bf 5}, 479 (1988)

\bibitem{V}
V. M. Vilalba, preprint gr-qc/9306019 


\bibitem{BJDR} 
J. D. Bjorken and S. D. Drell S.D.  {\it Relativistic Quantum Mechanics}, 
McGraw-Hill Book Co., NY, 1964

\bibitem{TH} 
B. Thaller,  {\it The Dirac Equation}, Springer Verlag, Berlin 
Heidelberg, 1992

\bibitem{BD}
N. D. Birrel and P. C. W. Davies, {\it Quantum Fields in Curved Space}, 
Cambridge University Press, Cambridge (1982)

\bibitem{SUP}
R. Dutt, A. Khare and U. P. Sukhatme, Am. J. Phys. {\bf 56}, 163 (1989);
F. Cooper,  A. Khare and U. P. Sukhatme, Phys. Rep.  {\bf 251}, 267 (1995)

\bibitem{AS}
M. Abramowitz and I. A. Stegun, {\it Handbook of Mathematical Functions}
(Dover, 1964)

 
\end{thebibliography}
\end{document}